\documentclass[aps,prb,longbibliography,reprint,superscriptaddress,citeautoscript]{revtex4-1}

\usepackage[applemac]{inputenc}
\usepackage[T1]{fontenc}
\usepackage[american]{babel}
\usepackage[scaled]{helvet}
\usepackage{graphicx,amssymb,bm,mathtools,textcomp,courier,multirow,bigdelim,color,hyperref}
\usepackage[normalem]{ulem}	%fixes problem with book references

\input{Qcircuit}
\graphicspath{{Pictures/}}

\begin{document}
\def\sectionautorefname{Sec.}

\title{Simple operation sequences to couple and interchange quantum information between spin qubits of different kinds}

\author{Sebastian Mehl}
\email{s.mehl@fz-juelich.de}
\author{David P. DiVincenzo}
\affiliation{JARA-Institute for Quantum Information, RWTH Aachen University, D-52056 Aachen, Germany}
\affiliation{Peter Gr\"unberg Institute (PGI-2), Forschungszentrum J\"ulich, D-52425 J\"ulich, Germany}

\date{\today}

%------------------------------------------------------------------------------------
%------------------------------------------------------------------------------------
%------------------------------------------------------------------------------------
\begin{abstract}
Efficient operation sequences to couple and interchange quantum information between quantum dot spin qubits of different kinds are derived using exchange interactions. In the qubit encoding of a single-spin qubit, a singlet-triplet qubit, and an exchange-only (triple-dot) qubit, some of the single-qubit interactions remain on during the entangling operation; this greatly simplifies the operation sequences that construct entangling operations. In the ideal setup, the gate operations use the intra-qubit exchange interactions only once. The limitations of the entangling sequences are discussed, and it is shown how quantum information can be converted between different kinds of quantum dot spin qubits.
\end{abstract}

\maketitle

%------------------------------------------------------------------------------------
%------------------------------------------------------------------------------------
%------------------------------------------------------------------------------------
\section{Introduction}

Small arrays of singly occupied quantum dot (QD) qubits are now fabricated in GaAs and silicon with great reliability.\cite{hanson2007-2,zwanenburg2013} These setups are of high interest for quantum computation because the electron spin can be used as a qubit.\cite{loss1998} Besides the single-spin qubit encoding, also more advanced qubit encodings have been suggested. Most promising are the singlet-triplet qubit (STQ)\cite{levy2002} and the exchange-only qubit.\cite{divincenzo2000} These qubits encode quantum information in the $s_z=0$ spin subspace of a two-electron double QD (DQD) or in two of the eight possible spin configurations of a three-electron triple QD (TQD).

For all the described qubits, single-qubit gates have been realized with high fidelities. Electric \cite{kawakami2014,scarlino2015} or magnetic \cite{pla2012,veldhorst2014,veldhorst2015} field pulses can nowadays control single spins with very high fidelities. High-fidelity gates for STQs are also possible when the electron configuration of the DQD is modified while the magnetic field across the DQD is inhomogeneous.\cite{hanson2007-1} Experimentally, a preparation of the nuclear magnetic field \cite{foletti2009,bluhm2010} or a micromagnet\cite{wu2014} created such static magnetic field configurations. The three-electron TQD can be operated using exchange interactions alone;\cite{divincenzo2000,medford2013-2,eng2015} more optimal qubit control has been realized if some of the exchange interactions are not reduced to zero.\cite{taylor2013,medford2013} Two-qubit gates have been proposed for all the qubit encodings using exchange couplings,\cite{loss1998,levy2002,mehl2014-1,divincenzo2000,doherty2013} while experiments have realized these gates only for single-spin qubits.\cite{veldhorst2014-2} STQs or exchange-only qubits can be coupled indirectly via their charge sector, e.g., using Coulomb interactions\cite{taylor2005} or couplings via cavity modes\cite{taylor2006-2,burkard2006,taylor2013}. These approaches have not been successful yet due to a high amount of dephasing that is caused by charge noise.\cite{shulman2014,frey2012,petersson2012,toida2013}

The present study assumes that universal qubit control is possible for the encoded qubits, while two QDs from different qubits are exchange coupled. Operation sequences for entanglement generation and qubit conversion are derived between QD qubits of different kinds. The operation sequences profit from always-on single qubit Hamiltonians during the entanglement sequences, as in earlier studies of TQDs.\cite{weinstein2005,doherty2013} For STQs, the magnetic fields at the QDs should be prepared independently. Their values need to differ anyway to realize single-qubit control. For the exchange-only qubit, a linear TQD arrangement is considered. Here, the exchange couplings between the neighboring pairs of QDs remain always at similar magnitudes. Such setups have been used in a previous experiment\cite{taylor2013,medford2013}, and can be realized if the occupation of the middle QD is made unfavorable by increasing its chemical potential compared to the outer QDs.\cite{taylor2013,medford2013}

The main finding of this paper are explicit operation sequences to entangle QD qubits of different kinds. The always-on single-qubit couplings greatly simplify the operation sequences because they reduce the possibility of leakage from the computational subspace. Effective Hamiltonians and entangling sequences are derived; the setups only require two operation sequences to entangle a single-spin qubit and a STQ (or an exchange-only qubit and a STQ), or four operation sequences to entangle a single-spin qubit and an exchange-only qubit. It is shown how the entanglement sequences can be used to swap quantum information between the qubits, and the limitations of the operation sequences are discussed.

The simplicity of the entangling operations shows that a large lattice of QD qubits does not necessarily need to contain identical types of coded qubits (cf., e.g., the description of large scale quantum computation with STQs in \rcite{mehl2015-2}). One can easily convert and couple different QD qubits using the operation sequences derived in this paper. As a consequence, it is possible to use a qubit encoding just for the situation when it is most optimal. It is known that single-spin qubits have exceptionally long coherence times, which makes them an ideal quantum memory.\cite{balasubramanian2009,pla2012} Encoded spin qubits, like the STQ or the exchange-only qubit, can be employed in their orbital sector, which makes them more ideal for readout or for long-distance couplings.\cite{taylor2005,hanson2007-2} It is also possible to use the described operation sequences to couple QD spin qubits to other spin qubits, like, e.g., donor-bound spin qubits.\cite{kane1998} The electron spin bound to a donor atom is a well-known qubit candidate with many impressive experiments of coherent spin control in recent years.\cite{pla2012,muhonen2014,dehollain2014,muhonen2015} Also tunnel couplings between donor-bound and gate-defined spin qubits were shown recently.\cite{urdampilleta2015,foote2015}

The organization of the paper is as follows. \sref{sec:Def} introduces the mathematical descriptions of the single-spin qubit, the STQ, and the exchange-only qubit. \sref{sec:Interface} derives the operation sequences to entangle QD qubits of different qubit encodings. \sref{sec:Discussion} discusses the limitations of these operations and describes how quantum information is converted between different qubits. Finally, the results of the paper are summarized.

%------------------------------------------------------------------------------------
%------------------------------------------------------------------------------------
%------------------------------------------------------------------------------------
\section{\label{sec:Def}
Qubit Definitions}

%------------------------------------------------------------------------------------
%------------------------------------------------------------------------------------
%------------------------------------------------------------------------------------
\subsection{Single-spin qubit}

A single spin defines a qubit using the states $\ket{0}=\ket{\uparrow}$ and $\ket{1}=\ket{\downarrow}$.\cite{loss1998} Universal qubit control is realized when a magnetic field can be tilted to two different directions. The control mechanisms to manipulate spins are magnetic field pulses\cite{koppens2006,veldhorst2014}, moving spins in static magnetic fields with spin-orbit interactions,\cite{nowack2007} and driving spins through areas of different magnetic fields.\cite{brunner2011,yoneda2014,kawakami2014} Without further discussing the exact mechanism, it is assumed here that the magnetic field direction can be rotated to the z and x directions to generate rotations around the z and x axes of the Bloch sphere. These single-qubit gates are labeled $\text{Z}_\phi=e^{-i2\pi\frac{\phi}{2}\sigma_z}$ and $\text{X}_\phi=e^{-i2\pi\frac{\phi}{2}\sigma_x}$, where $\sigma_z=\op{\uparrow}{\uparrow}-\op{\downarrow}{\downarrow}$ and $\sigma_x=\op{\uparrow}{\downarrow}+\op{\downarrow}{\uparrow}$ are the Pauli operators. The phase accumulation $\phi=\frac{E_z}{h}$ ($\phi=\frac{E_x}{h}$) is caused by the Zeeman energy $E_z=g\mu_B B_z$ ($E_x=g\mu_B B_x$) of the magnetic field in the z direction (x direction).\footnote{In contrast to \rcite{nielsen2000}, all the phase accumulations are given in multiples of $2\pi$.}

%------------------------------------------------------------------------------------
%------------------------------------------------------------------------------------
%------------------------------------------------------------------------------------
\subsection{Singlet-triplet qubit}

STQs are coded using the $s_z=0$ spin subspace of a two-electron DQD.\cite{levy2002} QD$_1$ and QD$_2$ label the individual QDs of the DQD. Ideally, the electrons are spatially separated, and each QD is occupied with one electron. The logical qubit states are defined by $\ket{0}=\ket{\uparrow\downarrow}$ and $\ket{1}=\ket{\downarrow\uparrow}$, where the first entry labels the electron at QD$_1$, and the second entry labels the electron at QD$_2$. Single-qubit control is realized using a magnetic field gradient between the QDs, corresponding to energy differences $\frac{\Delta E_z}{2}\left(\sigma_{z,1}-\sigma_{z,2}\right)$, with $\Delta E_z=(E_{z,1}-E_{z,2})/2$, and the exchange interaction between the QD electrons $\frac{J_{12}}{4}\left(\bm{\sigma}_1\cdot\bm{\sigma}_2-\bm{1}\right)$. $\bm{\sigma}_i=\left(\sigma_{x,i},\sigma_{y,i},\sigma_{z,i}\right)$ is the vector of Pauli matrices at QD$_i$.

$\Delta E_z$ is usually static in experiments, but $J_{12}$ can be tuned within subnanoseconds by controlling the tunnel coupling or the potential difference of the QDs.\cite{petta2005} The magnetic field gradient generates rotations around the z axis of the Bloch sphere $\text{Z}_\phi=e^{-i2\pi\frac{\phi}{4}\left(\sigma_{z,1}-\sigma_{z,2}\right)}$, with $\phi=2\Delta E_z/h$, and rotations around the x axis are caused by the exchange interaction $\text{X}_\phi=e^{-i2\pi\frac{\phi}{4}\left(\bm{\sigma}_1\cdot\bm{\sigma}_2-\bm{1}\right)}$, with $\phi=J_{12}/h$. To reduce the leakage probability, experiments are always done at global magnetic fields $\frac{E_z}{2}\left(\sigma_{z,1}+\sigma_{z,2}\right)$, with $E_z=(E_{z,1}+E_{z,2})/2$, that lift the degeneracy between the leakage states $\left\{\ket{\uparrow\uparrow},\ket{\downarrow\downarrow}\right\}$ and the computational subspace $\left\{\ket{0},\ket{1}\right\}$.

%------------------------------------------------------------------------------------
%------------------------------------------------------------------------------------
%------------------------------------------------------------------------------------
\subsection{Exchange-only qubit}

The exchange-only qubit is coded using the $S=\frac{1}{2}$, $s_z=\frac{1}{2}$ subspace of three electrons.\cite{divincenzo2000} The following operations require the initialization to a subspace encoding.\cite{mehl2013-1} In any case, applying strong, global external magnetic field eases controlled state initializations. The three singly-occupied QDs are labeled by QD$_1$, QD$_2$, and QD$_3$. The qubit states are defined by $\ket{0}=\frac{1}{\sqrt{2}}\left(\ket{\uparrow\uparrow\downarrow}-\ket{\downarrow\uparrow\uparrow}\right)$ and $\ket{1}=\frac{1}{\sqrt{6}}\left(\ket{\uparrow\uparrow\downarrow}+\ket{\downarrow\uparrow\uparrow}\right)-\sqrt{\frac{2}{3}}\ket{\uparrow\downarrow\uparrow}$, with the spin labels $\ket{\sigma_{\text{QD}_1},\sigma_{\text{QD}_2},\sigma_{\text{QD}_3}}$. The sum of the exchange interactions $\frac{J}{4}\left[
\left(\bm{\sigma_1}\cdot\bm{\sigma_2}-\bm{1}\right)+
\left(\bm{\sigma_2}\cdot\bm{\sigma_3}-\bm{1}\right)\right]$, with $J=\left(J_{12}+J_{23}\right)/2$, and their difference $\frac{\Delta J}{4}\left[
\left(\bm{\sigma_1}\cdot\bm{\sigma_2}-\bm{1}\right)-
\left(\bm{\sigma_2}\cdot\bm{\sigma_3}-\bm{1}\right)\right]$, with $\Delta J=\left(J_{12}-J_{23}\right)/2$, provide universal control of the subspace $\left\{\ket{0},\ket{1}\right\}$. $J$ causes a rotation around the z axis of the Bloch sphere $\text{Z}_{\phi}=e^{-i2\pi\frac{\phi}{4}\left[
\left(\bm{\sigma_1}\cdot\bm{\sigma_2}-\bm{1}\right)+
\left(\bm{\sigma_2}\cdot\bm{\sigma_3}-\bm{1}\right)\right]}$, with $\phi=J/h$, and $\Delta J$ causes a rotation around the x axis $\text{X}_{\phi}=e^{-i2\pi\frac{\phi}{4\sqrt{3}}\left[
\left(\bm{\sigma_1}\cdot\bm{\sigma_2}-\bm{1}\right)-
\left(\bm{\sigma_2}\cdot\bm{\sigma_3}-\bm{1}\right)\right]}$, with $\phi=\sqrt{3}\Delta J/h$. In typical qubit manipulation protocols, $J$ is constant and large, while $\Delta J$ is rapidly driven around zero.\cite{taylor2013,medford2013}

%------------------------------------------------------------------------------------
%------------------------------------------------------------------------------------
%------------------------------------------------------------------------------------
\section{\label{sec:Interface}
Interfaces between spin qubits}

%------------------------------------------------------------------------------------
%------------------------------------------------------------------------------------
%------------------------------------------------------------------------------------
\subsection{Single-spin qubit and singlet-triplet qubit}

\begin{figure}
\includegraphics[width=0.35\textwidth]{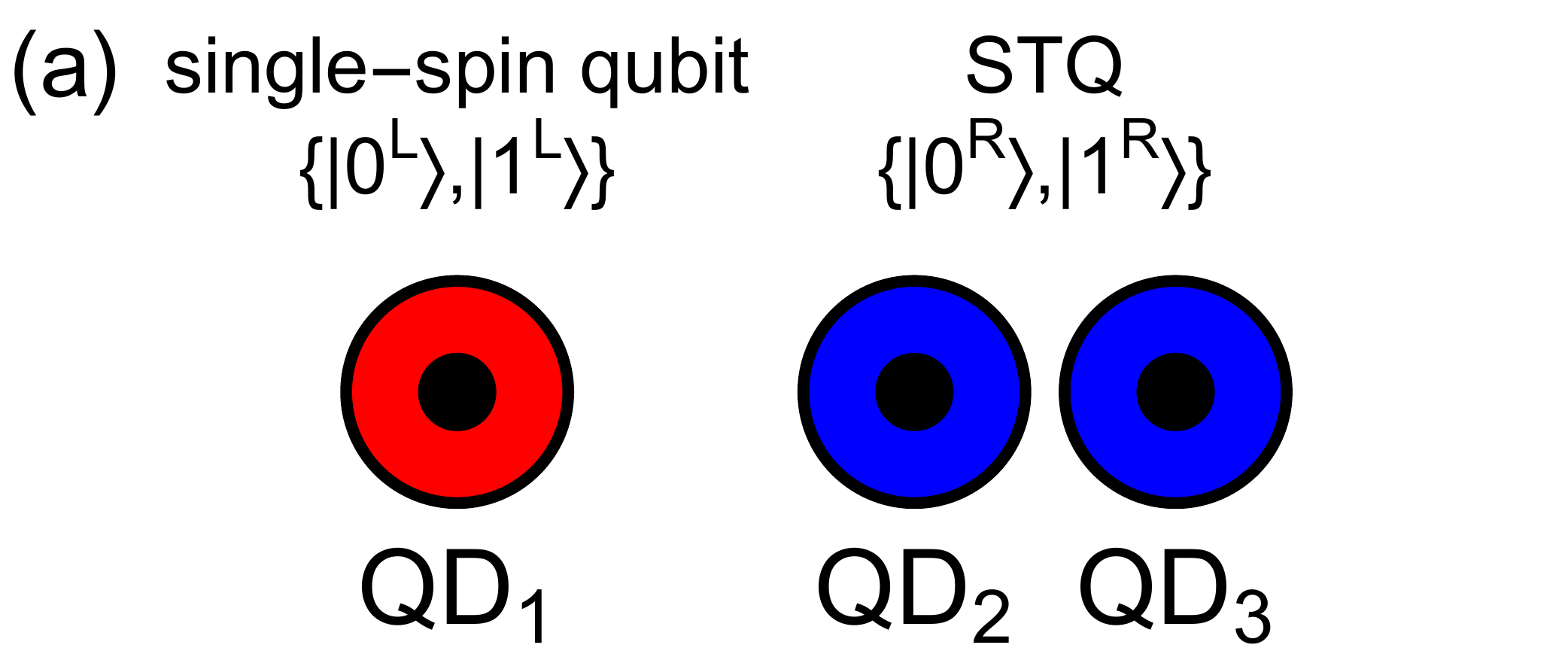}
\[
\Qcircuit @C=.5em @R=1em @!R {
\push{\text{(b)}\rule{2.75em}{0em}}&\lstick{\text{QD}_1}&\ctrl{1}&\qw&&&\multigate{2}{\mathcal{U}^{\text{A}}_{1/2,\sqrt{3}/2}}&\gate{\text{Z}_{(1-\sqrt{3})/4}^{\text{L}}}&\qw
\\
&\lstick{\text{QD}_2}&\multigate{1}{\text{Z}}&\qw& \push{\rule{.3em}{0em}=\rule{.3em}{0em}} &&\ghost{\mathcal{U}^{\text{A}}_{1/2,\sqrt{3}/2}}&\multigate{1}{\text{Z}_{(1-3\sqrt{3})/4}^{\text{R}}}&\qw
\\
&\lstick{\text{QD}_3}&\ghost{\text{Z}}&\qw&&&\ghost{\mathcal{U}^{\text{A}}_{1/2,\sqrt{3}/2}}&\ghost{\text{Z}_{(1-3\sqrt{3})/4}^{\text{R}}}&\qw
}
\]
\caption{\label{fig:01}
Entangling operation between a single-spin qubit and a STQ. (a) QD$_1$ defines a single-spin qubit with the qubit levels $\{\ket{0^{\text{L}}},\ket{1^{\text{L}}}\}$; QD$_2$ and QD$_3$ define a STQ with the qubit levels $\{\ket{0^{\text{R}}},\ket{1^{\text{R}}}\}$. A weak tunnel coupling between QD$_1$ and QD$_2$ couples the single-spin qubit and the STQ. (b) Sequence to create a CPHASE between a single-spin qubit (coded on QD$_1$) and a STQ (coded on QD$_2$ and QD$_3$). $\text{Z}^{\text{L}}_{\phi}$ and $\text{Z}^{\text{R}}_{\phi}$ are the phase gates of the qubits L and R. $\mathcal{U}_{\phi,\psi}^{\text{A}}$ is defined in \eref{eq:Unitary_LDSTQ}.
}
\end{figure}

\fref{fig:01}(a) shows a trio of singly-occupied QDs that encodes a single-spin qubit and a STQ. QD$_1$ defines the single-spin qubit, with the qubit levels $\{\ket{0^{\text{L}}},\ket{1^{\text{L}}}\}$. QD$_2$ and QD$_3$ define the STQ, where the qubit levels are called $\{\ket{0^{\text{R}}},\ket{1^{\text{R}}}\}$. A general Hamiltonian in this setup is
\begin{align}
\label{eq:Ham_LDSTQ}
\mathcal{H}^{\text{A}}=&
\frac{J_{12}}{4}\left(\bm{\sigma_1}\cdot\bm{\sigma_2}-\bm{1}\right)+
\frac{E_z}{2}\left(\sigma_{z,1}+\sigma_{z,2}+\sigma_{z,3}\right)\\
&+
\frac{\widetilde{E}_{z,2}}{2}\sigma_{z,2}+
\frac{\widetilde{E}_{z,3}}{2}\sigma_{z,3}.
\nonumber
\end{align}
QD$_1$ and QD$_2$ are coupled by the exchange coupling $J_{12}$ that is described by the first term in \eref{eq:Ham_LDSTQ}. The second term describes the global magnetic field $E_z$, and the last two terms are the deviations of the local magnetic fields at QD$_2$ and QD$_3$ from $E_z$. To construct entangling operations, $\frac{E_z}{2}\left(\sigma_{z,1}+\sigma_{z,2}+\sigma_{z,3}\right)$ and $\frac{\widetilde{E}_{z,3}}{2}\sigma_{z,3}$ can be neglected because these terms commute with the remaining parts of \eref{eq:Ham_LDSTQ}, and they generate only irrelevant phases. The relevant time evolution is described by
\begin{align}
\mathcal{U}_{\phi,\psi}^{\text{A}}=e^{-i2\pi\left[
\frac{\phi}{4}\left(\bm{\sigma_1}\cdot\bm{\sigma_2}-\bm{1}\right)+
\frac{\psi}{2}\sigma_{z,2}\right]},
\label{eq:Unitary_LDSTQ}
\end{align}
with $\phi=J_{12}/h$ and $\psi=\widetilde{E}_{z,2}/h$.

Only the states in the subspace $\{\ket{0^{\text{L}}0^{\text{R}}},$$\ket{0^{\text{L}}1^{\text{R}}},$$\ket{1^{\text{L}}0^{\text{R}}},$ $\ket{1^{\text{L}}1^{\text{R}}},$$\ket{\downarrow\uparrow\uparrow},$$\ket{\uparrow\downarrow\downarrow}\}$ are coupled in \eref{eq:Unitary_LDSTQ}. There is no evolution from computational states to leakage states for $\sqrt{\phi^2+\psi^2}=\mathbb{Z}$. An entangling operation that is, up to local unitaries, equivalent to the CPHASE operation is realized for $\phi=\mathbb{Z}+\frac{1}{2}$. One can used, e.g., $\mathcal{U}_{1/2,\sqrt{3}/2}^{\text{A}}$. A CPHASE in the basis 
$\left\{\ket{0^{\text{L}}0^{\text{R}}},\ket{0^{\text{L}}1^{\text{R}}},\ket{1^{\text{L}}0^{\text{R}}},\ket{1^{\text{L}}1^{\text{R}}}\right\}$ is [see \fref{fig:01}(b)]:
\begin{align}
\text{Z}_{(1-\sqrt{3})/4}^{\text{L}}
\text{Z}_{(1-3\sqrt{3})/4}^{\text{R}}
\mathcal{U}_{1/2,\sqrt{3}/2}^{\text{A}}=e^{-i\frac{\pi}{2}}\text{CPHASE}.
\end{align}

%------------------------------------------------------------------------------------
%------------------------------------------------------------------------------------
%------------------------------------------------------------------------------------
\subsection{\label{sec:SSRX}
Single-spin qubit and exchange-only qubit}

\begin{figure}
\includegraphics[width=0.35\textwidth]{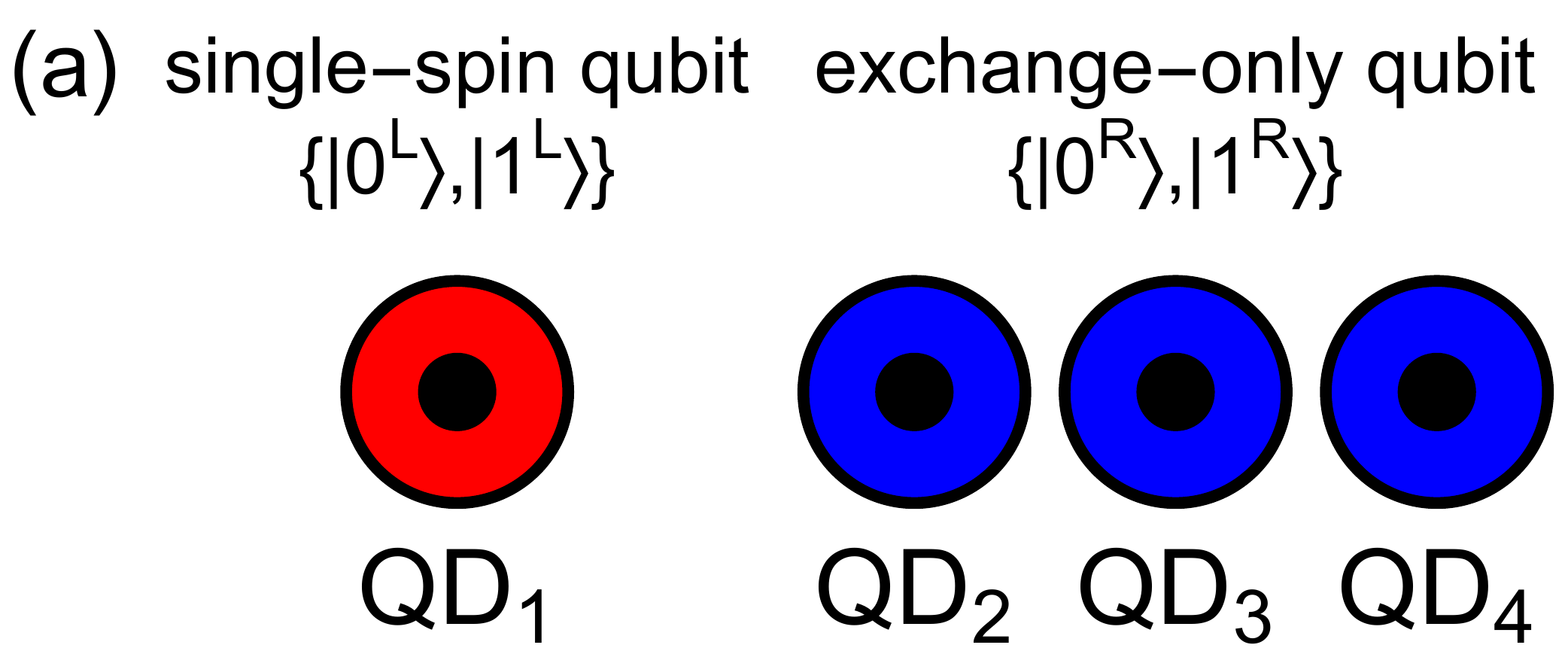}
\[
\Qcircuit @C=.5em @R=1em @!R {
\push{\text{(b)}\rule{2.75em}{0em}}&\lstick{\text{QD}_1}&\ctrl{1}&\qw&&&\multigate{3}{\mathcal{U}^{\text{B}}_{3/4}}&\gate{\text{Z}_{1/2}^{\text{L}}}&\multigate{3}{\mathcal{U}^{\text{B}}_{3/4}}&\gate{\text{Z}_{1/2}^{\text{L}}}&\qw
\\
&\lstick{\text{QD}_2}&\multigate{2}{\text{Z}}&\qw& \push{\rule{.3em}{0em}=\rule{.3em}{0em}} &&\ghost{\mathcal{U}^{\text{B}}_{3/4}}&\qw&\ghost{\mathcal{U}^{\text{B}}_{3/4}}&\multigate{2}{\text{Z}_{1/4}^{\text{R}}}&\qw
\\
&\lstick{\text{QD}_3}&\ghost{\text{Z}}&\qw&&&\ghost{\mathcal{U}^{\text{B}}_{3/4}}&\qw&\ghost{\mathcal{U}^{\text{B}}_{3/4}}&\ghost{\text{Z}_{1/4}^{\text{R}}}&\qw
\\
&\lstick{\text{QD}_4}&\ghost{\text{Z}}&\qw&&&\ghost{\mathcal{U}^{\text{B}}_{3/4}}&\qw&\ghost{\mathcal{U}^{\text{B}}_{3/4}}&\ghost{\text{Z}_{1/4}^{\text{R}}}&\qw
}
\]
\caption{\label{fig:02}
Entangling operation between a single-spin qubit and an exchange-only qubit. (a) QD$_1$ defines a single-spin qubit with the qubit levels $\{\ket{0^{\text{L}}},\ket{1^{\text{L}}}\}$; QD$_2$-QD$_4$ define an exchange-only qubit with the qubit levels $\{\ket{0^{\text{R}}},\ket{1^{\text{R}}}\}$. A weak tunnel coupling between QD$_1$ and QD$_2$ couples the single-spin qubit and the exchange-only qubit. (b) Sequence to create a CPHASE between a single-spin qubit (coded on QD$_1$) and an exchange-only qubit (coded on QD$_2$-QD$_4$). $\text{Z}^{\text{L}}_{\phi}$ and $\text{Z}^{\text{R}}_{\phi}$ are the phase gates of the qubits L and R. $\mathcal{U}_{\phi}^{\text{B}}$ is defined in \eref{eq:HEff_2}.}
\end{figure}

\fref{fig:02}(a) shows a quartet of singly-occupied QDs that encodes a single-spin qubit (QD$_1$; qubit states $\{\ket{0^{\text{L}}},\ket{1^{\text{L}}}\}$) and an exchange-only qubit (QD$_2$-QD$_4$; qubit states $\{\ket{0^{\text{R}}},\ket{1^{\text{R}}}\}$). A general interaction in this setup is
\begin{align}
\nonumber
\mathcal{H}^{\text{B}}=&
\frac{J_{12}}{4}\left(\bm{\sigma_1}\cdot\bm{\sigma_2}-\bm{1}\right)
+
\frac{E_z}{2}\left(\sigma_{z,1}+\sigma_{z,2}+\sigma_{z,3}+\sigma_{z,4}\right)\\
&+\frac{J}{4}\left[
\left(\bm{\sigma_2}\cdot\bm{\sigma_3}-\bm{1}\right)+
\left(\bm{\sigma_3}\cdot\bm{\sigma_4}-\bm{1}\right)\right].
\label{eq:Ham_LDRX}
\end{align}
The first term in \eref{eq:Ham_LDRX} is the exchange coupling between QD$_1$ and QD$_2$. The second term is the global magnetic field, and the third term describes the exchange couplings of the exchange-only qubit. 

$\frac{E_z}{2}\left(\sigma_{z,1}+\sigma_{z,2}+\sigma_{z,3}+\sigma_{z,4}\right)$ commutes with the remaining parts of \eref{eq:Ham_LDRX}, and this term causes only an irrelevant time evolution of the single-spin qubit. The relevant time evolution through \eref{eq:Ham_LDRX} is
\begin{align}
\mathcal{U}_{\phi,\psi}^{\text{B}}=
e^{-i2\pi\left\{
\frac{\phi}{4}\left(\bm{\sigma_1}\cdot\bm{\sigma_2}-\bm{1}\right)+
\frac{\psi}{4}
\left[
\left(\bm{\sigma_2}\cdot\bm{\sigma_3}-\bm{1}\right)+\left(\bm{\sigma_3}\cdot\bm{\sigma_4}-\bm{1}\right)
\right]
\right\}},
\label{eq:Unitary_LDRX}
\end{align}
with $\phi=J_{12}/h$ and $\psi=J/h$. There are exact entangling operations between a single-spin qubit and an exchange-only qubit that use \eref{eq:Unitary_LDRX}. However, these sequences are complicated and involve many operation steps.\footnote{
I found operation sequences to create entangling operations with a numerical search algorithm, similar to the description in \rcite{mehl2014-1}. An operation sequence that is equivalent to a CPHASE is
\begin{align*}
\mathcal{U}_{\phi_1,\phi_2}^{\text{B}}
\text{X}_{\phi_3}^{\text{R}}
\mathcal{U}_{\phi_1,\phi_2}^{\text{B}}
\text{Z}_{1/2}^{\text{L}}
\text{X}_{\phi_4}^{\text{R}}
\mathcal{U}_{\phi_1,\phi_2}^{\text{B}}
\text{X}_{\phi_3}^{\text{R}}
\mathcal{U}_{\phi_1,\phi_2}^{\text{B}},
\end{align*}
with
$\phi_1=0.195613200942698$,
$\phi_2=0.2178346646839128$,
$\phi_3=0.7362256575556158$, and
$\phi_4=0.735072280195903$.}

Simpler entangling operations can be constructed for $J\gg J_{12}$. The computational subspace is part of the four-spin subspaces $S=0$, $s_z=0$ and $S=1$, $s_z=1,0$, which together have eight dimensions.\cite{[{See the standard spin addition rules, e.g., in }]{sakurai1994}} Because the Hamiltonian in \eref{eq:Ham_LDRX} preserves the spin quantum numbers, it is sufficient to describe the time evolution only in the four-spin subspaces $S=0$, $s_z=0$ and $S=1$, $s_z=1,0$ that are spanned by $\left\{\ket{0^{\text{L}}0^{\text{R}}},\ket{0^{\text{L}}1^{\text{R}}},\ket{1^{\text{L}}0^{\text{R}}},\ket{l_1},\ket{1^{\text{L}}1^{\text{R}}},\ket{l_2},\ket{l_3},\ket{l_4}\right\}$, with 
$\ket{l_1}=\ket{0^{\text{L}}}\ket{u_{-1/2}}$, 
$\ket{l_2}=\ket{0^{\text{L}}}\ket{v_{-1/2}}$,
$\ket{l}_3\propto\ket{0^{\text{L}}}\ket{Q_{-1/2}}-\ket{1^{\text{L}}}\ket{Q_{1/2}}$, and
$\ket{l}_4\propto\sqrt{3}\ket{1^{\text{L}}}\ket{Q_{3/2}}-\ket{0^{\text{L}}}\ket{Q_{1/2}}$. The states $\ket{u_{-1/2}}=\frac{1}{\sqrt{2}}\left(\ket{\uparrow\downarrow\downarrow}-\ket{\downarrow\downarrow\uparrow}\right)$ and $\ket{v_{-1/2}}=\frac{1}{\sqrt{6}}\left(\ket{\uparrow\downarrow\downarrow}+\ket{\downarrow\downarrow\uparrow}\right)-\sqrt{\frac{2}{3}}\ket{\downarrow\uparrow\downarrow}$ span the $S=\frac{1}{2}$, $s_z=-\frac{1}{2}$ spin subspace of three electrons; $\ket{Q_{3/2}}=\ket{\uparrow\uparrow\uparrow}$, $\ket{Q_{1/2}}\propto\ket{\uparrow\uparrow\downarrow}+\ket{\uparrow\downarrow\uparrow}+\ket{\downarrow\uparrow\uparrow}$, $\ket{Q_{-1/2}}\propto\ket{\downarrow\downarrow\uparrow}+\ket{\downarrow\uparrow\downarrow}+\ket{\uparrow\downarrow\downarrow}$, and $\ket{Q_{-3/2}}=\ket{\downarrow\downarrow\downarrow}$ are the $S=\frac{3}{2}$ quadruplet states of three spins. The spin labels correspond to $\ket{\sigma_{\text{QD}_2},\sigma_{\text{QD}_3},\sigma_{\text{QD}_4}}$ in these state definitions.

The projection of \eref{eq:Ham_LDRX} to the given basis is
\begin{widetext}
\begin{align}
\mathcal{H}^{\text{B}}=\left(\begin{array}{cccccccc}
%%%%%%%%%%%%%%%%%%%%%%%%%%%%%%%%%%%%%%
\multicolumn{1}{c|}{E_z-\frac{J}{2}-\frac{J_{12}}{4}}&
\frac{J_{12}}{4\sqrt{3}}&
&
&
&
&
&
-\frac{J_{12}}{\sqrt{6}}
\\\cline{1-2}
%%%%%%%%%%%%%%%%%%%%%%%%%%%%%%%%%%%%%%
\frac{J_{12}}{4\sqrt{3}}&
\multicolumn{1}{|c|}{E_z-\frac{3J}{2}-\frac{J_{12}}{12}}&
&
&
&
&
&
\frac{J_{12}}{3\sqrt{2}}
\\\cline{2-4}
%%%%%%%%%%%%%%%%%%%%%%%%%%%%%%%%%%%%%%
&
&
\multicolumn{1}{|c}{-\frac{J}{2}-\frac{J_{12}}{4}}&
\multicolumn{1}{c|}{0}&
-\frac{J_{12}}{4\sqrt{3}}&
-\frac{J_{12}}{2\sqrt{3}}&
\frac{J_{12}}{2\sqrt{3}}&\\
%%%%%%%%%%%%%%%%%%%%%%%%%%%%%%%%%%%%%%
&
&
\multicolumn{1}{|c}{0}&
\multicolumn{1}{c|}{-\frac{J}{2}-\frac{J_{12}}{4}}&
\frac{J_{12}}{2\sqrt{3}}&
\frac{J_{12}}{4\sqrt{3}}&
\frac{J_{12}}{2\sqrt{3}}&
\\\cline{3-6}
%%%%%%%%%%%%%%%%%%%%%%%%%%%%%%%%%%%%%%
&
&
-\frac{J_{12}}{4\sqrt{3}}&
\frac{J_{12}}{2\sqrt{3}}&
\multicolumn{1}{|c}{-\frac{3J}{2}-\frac{5J_{12}}{12}}&
-\frac{J_{12}}{3}&
\multicolumn{1}{|c}{-\frac{J_{12}}{6}}&\\
%%%%%%%%%%%%%%%%%%%%%%%%%%%%%%%%%%%%%%
&
&
-\frac{J_{12}}{2\sqrt{3}}&
\frac{J_{12}}{4\sqrt{3}}&
\multicolumn{1}{|c}{-\frac{J_{12}}{3}}&
-\frac{3J}{2}-\frac{5J_{12}}{12}&
\multicolumn{1}{|c}{\frac{J_{12}}{6}}&
\\\cline{5-7}
%%%%%%%%%%%%%%%%%%%%%%%%%%%%%%%%%%%%%%
&
&
\frac{J_{12}}{2\sqrt{3}}&
\frac{J_{12}}{2\sqrt{3}}&
-\frac{J_{12}}{6}&
\frac{J_{12}}{6}&
\multicolumn{1}{|c|}{-\frac{2J_{12}}{3}}&
\\\cline{7-8}
%%%%%%%%%%%%%%%%%%%%%%%%%%%%%%%%%%%%%%
-\frac{J_{12}}{\sqrt{6}}&
\frac{J_{12}}{3\sqrt{2}}&
&
&
&
&
\multicolumn{1}{c|}{}&
E_z-\frac{2J_{12}}{3}
%%%%%%%%%%%%%%%%%%%%%%%%%%%%%%%%%%%%%%
\end{array}\right).
\label{eq:HEff}
\end{align}
\end{widetext}
It is sufficient to consider the time evolution in the subspaces of equal energies that are defined by $E_z$ and $J$. The borders in the matrix of \eref{eq:HEff} indicate these subspaces. $J_{12}$ couples these subspaces, but for $E_z,J\gg J_{12}$ these processes can be neglected.

After neglecting all the entries outside of the marked subspaces in \eref{eq:HEff}, also the time evolutions of $E_z$ and $J$ factor because they commute with the remaining entries. The global magnetic field
$\frac{E_z}{2}(\sigma_{z,1}+\sigma_{z,2}+\sigma_{z,3}+\sigma_{z,4})\simeq 
\frac{E_z}{2}
\left(\op{0^{\text{L}}}{0^{\text{L}}}-\op{1^{\text{L}}}{1^{\text{L}}}\right)$ and the exchange interaction
$\frac{J}{4}\left[
\left(\bm{\sigma_2}\cdot\bm{\sigma_3}-\bm{1}\right)+
\left(\bm{\sigma_3}\cdot\bm{\sigma_4}-\bm{1}\right)\right]\simeq
\frac{J}{2}
\left(\op{0^{\text{R}}}{0^{\text{R}}}-\op{1^{\text{R}}}{1^{\text{R}}}\right)
$ cause single-qubit time evolutions that will be neglected in the following. \eref{eq:Unitary_LDRX} can then be simplified on the subspace $\left\{\ket{0^{\text{L}}0^{\text{R}}},\ket{0^{\text{L}}1^{\text{R}}},\ket{1^{\text{L}}0^{\text{R}}},\ket{l_1},\ket{1^{\text{L}}1^{\text{R}}},\ket{l_2}\right\}$ to
\begin{align}
\mathcal{U}_\phi^{\text{B}}\approx e^{-i2\pi\phi ~\text{diag}\left\{
-\frac{1}{4},
-\frac{1}{12},
\left(\begin{array}{cc}
-\frac{1}{4}&0\\
0&-\frac{1}{4}
\end{array}\right),
\left(\begin{array}{cc}
-\frac{5}{12}&-\frac{1}{3}\\
-\frac{1}{3}&-\frac{5}{12}
\end{array}\right)
\right\}
}.
\label{eq:HEff_2}
\end{align}
$\text{diag}\left\{a,b,\dots\right\}$ describes the matrix with the diagonal entries $a$, $b$, $\dots$, and $\phi=J_{12}/h$.

A single time evolution under \eref{eq:HEff_2} is never entangling because the criterion to prevent leakage only permits single-qubit gates. The two-step sequence $\mathcal{U}_\phi^{\text{B}} Z_{1/2}^{\text{L}}\mathcal{U}_\phi^{\text{B}}$ is equivalent to a CPHASE gate for $\phi=\frac{3}{4}+\frac{3}{2}\mathbb{Z}$. A CPHASE operation in the basis $\left\{\ket{0^{\text{L}}0^{\text{R}}},\ket{0^{\text{L}}1^{\text{R}}},\ket{1^{\text{L}}0^{\text{R}}},\ket{1^{\text{L}}1^{\text{R}}}\right\}$ is created by [see \fref{fig:02}(b)]:
\begin{align}
\label{eq:CPHASE_B}
\text{Z}_{1/2}^{\text{L}}
\text{Z}_{1/4}^{\text{R}}
\mathcal{U}_{3/4}^{\text{B}}
\text{Z}_{1/2}^{\text{L}}
\mathcal{U}_{3/4}^{\text{B}}=
\text{CPHASE}.
\end{align}
Note that the implicit single-qubit phase evolutions through $E_z$ and $J$, that are neglected in \eref{eq:HEff_2}, need to be included in $\text{Z}_{1/2}^{\text{L}}$ and $\text{Z}_{1/4}^{\text{R}}$.

%------------------------------------------------------------------------------------
%------------------------------------------------------------------------------------
%------------------------------------------------------------------------------------
\subsection{\label{sec:STQRX}
Singlet-triplet qubit and exchange-only qubit}

\begin{figure}
\includegraphics[width=0.35\textwidth]{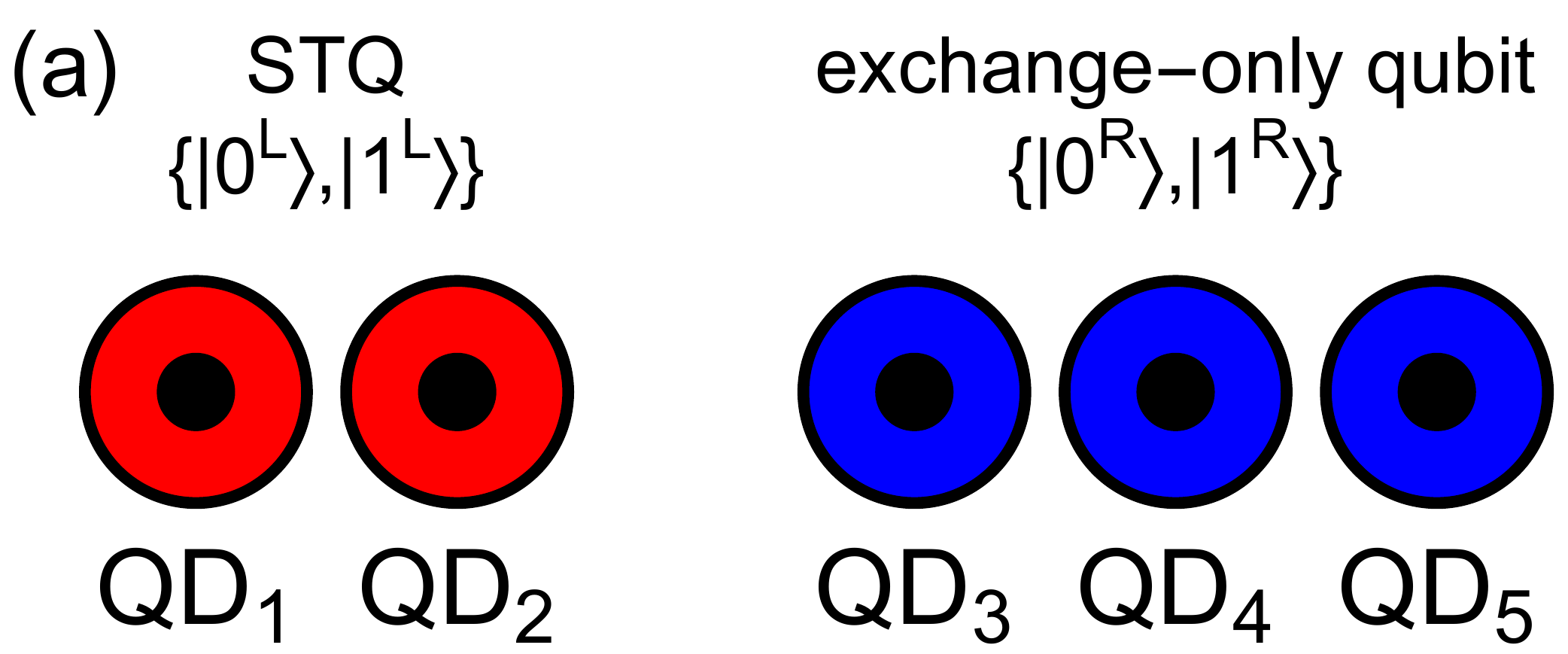}
\bigskip
\[
\Qcircuit @C=.5em @R=1em @!R {
\push{\text{(b)}\rule{2.75em}{0em}}&\lstick{\text{QD}_1}&\multigate{4}{\text{CZ}}&\qw&&&\multigate{4}{\mathcal{U}^{\text{C}_1}_{3/(2\sqrt{2}),1/\sqrt{2}}}&\multigate{1}{\text{Z}_{1/\sqrt{2}}^{\text{L}}}&\qw\\
&\lstick{\text{QD}_2}&\ghost{\text{CZ}}&\qw&&&\ghost{\mathcal{U}^{\text{C}_1}_{3/(2\sqrt{2}),1/\sqrt{2}}}&\ghost{\text{Z}_{1/\sqrt{2}}^{\text{L}}}&\qw\\
&\lstick{\text{QD}_3}&\ghost{\text{CZ}}&\qw&\push{\rule{.3em}{0em}=\rule{.3em}{0em}}&&\ghost{\mathcal{U}^{\text{C}_1}_{3/(2\sqrt{2}),1/\sqrt{2}}}&\multigate{2}{\text{Z}_{(4+\sqrt{2})/8}^{\text{R}}}&\qw\\
&\lstick{\text{QD}_4}&\ghost{\text{CZ}}&\qw&&&\ghost{\mathcal{U}^{\text{C}_1}_{3/(2\sqrt{2}),1/\sqrt{2}}}&\ghost{\text{Z}_{(3+\sqrt{2})/8}^{\text{R}}}&\qw\\
&\lstick{\text{QD}_5}&\ghost{\text{CZ}}&\qw&&&\ghost{\mathcal{U}^{\text{C}_1}_{3/(2\sqrt{2}),1/\sqrt{2}}}&\ghost{\text{Z}_{(3+\sqrt{2})/8}^{\text{R}}}&\qw
}
\]
\bigskip
\[
\Qcircuit @C=.5em @R=1em @!R {
\push{\text{(c)}\rule{2.75em}{0em}}&\lstick{\text{QD}_1}&\multigate{4}{\text{CZ}}&\qw&&&\multigate{1}{\text{H}^{\text{L}}}&\multigate{4}{\mathcal{U}^{\text{C}_2}_{3/\sqrt{2},1/(2\sqrt{2})}}&\multigate{1}{\text{H}^{\text{L}}}&\qw\\
&\lstick{\text{QD}_2}&\ghost{\text{CZ}}&\qw&&&\ghost{\text{H}^{\text{L}}}&\ghost{\mathcal{U}^{\text{C}_2}_{3/\sqrt{2},1/(2\sqrt{2})}}&\ghost{\text{H}^{\text{L}}}&\qw\\
&\lstick{\text{QD}_3}&\ghost{\text{CZ}}&\qw&\push{\rule{.3em}{0em}=\rule{.3em}{0em}}&&\qw&\ghost{\mathcal{U}^{\text{C}_2}_{3/\sqrt{2},1/(2\sqrt{2})}}&\multigate{2}{\text{Z}_{1/2}^{\text{R}}}&\qw\\
&\lstick{\text{QD}_4}&\ghost{\text{CZ}}&\qw&&&\qw&\ghost{\mathcal{U}^{\text{C}_2}_{3/\sqrt{2},1/(2\sqrt{2})}}&\ghost{\text{Z}_{1/2}^{\text{R}}}&\qw\\
&\lstick{\text{QD}_5}&\ghost{\text{CZ}}&\qw&&&\qw&\ghost{\mathcal{U}^{\text{C}_2}_{3/\sqrt{2},1/(2\sqrt{2})}}&\ghost{\text{Z}_{1/2}^{\text{R}}}&\qw
}
\]
\caption{\label{fig:03}
Entangling operations between a STQ and an exchange-only qubit. QD$_1$ and QD$_2$ define a STQ with the qubit levels $\{\ket{0^{\text{L}}},\ket{1^{\text{L}}}\}$; QD$_3$-QD$_5$ define an exchange-only qubit with the qubit levels $\{\ket{0^{\text{R}}},\ket{1^{\text{R}}}\}$. A weak tunnel coupling between QD$_2$ and QD$_3$ couples the STQ and the exchange-only qubit. (b)(c) Sequences to create a CPHASE between a STQ (coded on QD$_1$ and QD$_2$) and an exchange-only qubit (coded on QD$_3$-QD$_5$). $\text{Z}^{\text{L}}_{\phi}$ and $\text{Z}^{\text{R}}_{\phi}$ are the phase gates of the qubits L and R. $\mathcal{U}_{\phi,\psi}^{\text{C}_{1}}$ and $\mathcal{U}_{\phi,\psi}^{\text{C}_{2}}$ are defined in \eref{eq:HEff_4} and \eref{eq:HEff_6}. The CPHASE gate is abbreviated as CZ, and H$^{\text{L}}$ is the Hadamard gate for qubit L.}
\end{figure}

A quintet of singly-occupied QDs, as shown in \fref{fig:03}(a), defines a STQ (QD$_1$-QD$_2$; qubit states $\{\ket{0^{\text{L}}},\ket{1^{\text{L}}}\}$) and an exchange-only qubit (QD$_3$-QD$_5$; qubit states $\{\ket{0^{\text{R}}},\ket{1^{\text{R}}}\}$). A possible interaction in this setup is

\begin{align}
\label{eq:Ham_STQRX}
\mathcal{H}^{\text{C}_1}=&
\frac{J}{4}\left[
\left(\bm{\sigma_3}\cdot\bm{\sigma_4}-\bm{1}\right)+
\left(\bm{\sigma_4}\cdot\bm{\sigma_5}-\bm{1}\right)\right]\\\nonumber
&+\frac{J_{23}}{4}\left(\bm{\sigma}_2\cdot\bm{\sigma}_3-\bm{1}\right)
+\frac{\widetilde{E}_{z,2}}{2}\sigma_{z,2}\\\nonumber
&+\frac{E_z}{2}\left(\sigma_{z,1}+\sigma_{z,2}+\sigma_{z,3}+\sigma_{z,4}+\sigma_{z,5}\right).
\end{align}
The first term in \eref{eq:Ham_STQRX} describes the single-qubit interaction of the exchange-only qubit for $J_{34}=J_{45}$, with the abbreviation $J=\left(J_{34}+J_{45}\right)/2$. The second term is the exchange interaction between QD$_2$ and QD$_3$. A global magnetic field across all five QDs, $E_z$, is represented by the last term. $\widetilde{E}_{z,2}$ is a small deviation of the local magnetic field at QD$_2$ from the global magnetic field. Note that a possible deviation of the magnetic field at QD$_1$, $\widetilde{E}_{z,1}$, is irrelevant when the exchange interaction between QD$_1$ and QD$_2$ is reduced to zero. $\widetilde{E}_{z,1}$ would only cause single-qubit evolutions of the STQ.

The time evolution under \eref{eq:Ham_STQRX} can be used to construct an entangling operation between the STQ and the exchange-only qubit. Similar to the discussion in the previous section, $E_z$ and $J$ are much larger than $\widetilde{E}_{z,2}$ and $J_{23}$. Therefore the qubit time evolution can be described using only the five-spin subspaces $S=\frac{1}{2}$, $s_z=\frac{1}{2}$ and $S=\frac{3}{2}$, $s_z=\frac{1}{2}$ that have together nine dimensions.

For $E_z, J\gg \widetilde{E}_{z,2}, J_{23}$, only the states $\ket{m_1}=\ket{T_+}\ket{u_{-1/2}}$ and $\ket{m_2}=\ket{T_+}\ket{v_{-1/2}}$ coupled significantly to the computational subspace through \eref{eq:Ham_STQRX}. These states are eigenstates of $\frac{J}{4}\left[
\left(\bm{\sigma_3}\cdot\bm{\sigma_4}-\bm{1}\right)+
\left(\bm{\sigma_4}\cdot\bm{\sigma_5}-\bm{1}\right)\right]$, and they have identical energies as the qubit states. $\ket{u_{-1/2}}$ and $\ket{v_{-1/2}}$ span the $S=\frac{1}{2}$, $s_z=-\frac{1}{2}$ subspace of the spins at QD$_2$-QD$_4$ (using the definitions from \sref{sec:SSRX}). $m_3=\sqrt{\frac{1}{2}}\ket{T_-}\ket{Q_{3/2}}-\sqrt{\frac{1}{3}}\ket{T_0}\ket{Q_{1/2}}+\sqrt{\frac{1}{6}}\ket{T_+}\ket{Q_{-1/2}}$, $m_4=\sqrt{\frac{2}{5}}\ket{T_-}\ket{Q_{3/2}}+\sqrt{\frac{1}{15}}\ket{T_0}\ket{Q_{1/2}}-\sqrt{\frac{8}{15}}\ket{T_+}\ket{Q_{-1/2}}$, and $m_5=\ket{S}\ket{Q_{3/2}}$ have different energies, and therefore these states can be neglected. $\ket{T_+}=\ket{\uparrow\uparrow}$, $\ket{T_0}\propto\ket{\uparrow\downarrow}+\ket{\downarrow\uparrow}$, $\ket{T_-}=\ket{\downarrow\downarrow}$, and $\ket{S_0}=\ket{\uparrow\downarrow}-\ket{\downarrow\uparrow}$ are the usual triplet and singlet states at QD$_1$-QD$_2$. Projecting \eref{eq:Ham_STQRX} to $\left\{\ket{0^{\text{L}}0^{\text{R}}},\ket{1^{\text{L}}0^{\text{R}}},\ket{m_1},\ket{0^{\text{L}}1^{\text{R}}},\ket{1^{\text{L}}1^{\text{R}}},\ket{m_2}
\right\}$ gives

\begin{widetext}
\begin{align}
\mathcal{H}^{\text{C}_1}\approx\left(\begin{array}{cccccc}
%%%%%%%%%%%%%%%%%%%%%%%%%%%%%%%%%%%%%%
\frac{E_z-J-\widetilde{E}_{z,2}}{2}-\frac{J_{23}}{4}&
0&
\multicolumn{1}{c|}{0}&
-\frac{J_{23}}{4\sqrt{3}}&
&
-\frac{J_{23}}{2\sqrt{3}}\\
%%%%%%%%%%%%%%%%%%%%%%%%%%%%%%%%%%%%%%
0&
\frac{E_z-J+\widetilde{E}_{z,2}}{2}-\frac{J_{23}}{4}&
\multicolumn{1}{c|}{0}&
&
\frac{J_{23}}{4\sqrt{3}}&
\\
%%%%%%%%%%%%%%%%%%%%%%%%%%%%%%%%%%%%%%
0&
0&
\multicolumn{1}{c|}{\frac{E_z-J+\widetilde{E}_{z,2}}{2}-\frac{J_{23}}{4}}&
\frac{J_{23}}{2\sqrt{3}}&
&
\frac{J_{23}}{4\sqrt{3}}\\\cline{1-6}
%%%%%%%%%%%%%%%%%%%%%%%%%%%%%%%%%%%%%%
-\frac{J_{23}}{4\sqrt{3}}&
&
\multicolumn{1}{c|}{\frac{J_{23}}{2\sqrt{3}}}&
\frac{E_z-3J-\widetilde{E}_{z,2}}{2}-\frac{5J_{23}}{12}&
0&
-\frac{J_{23}}{3}\\
%%%%%%%%%%%%%%%%%%%%%%%%%%%%%%%%%%%%%%
&
\frac{J_{23}}{4\sqrt{3}}&
\multicolumn{1}{c|}{}&
0&
\frac{E_z-3J+\widetilde{E}_{z,2}}{2}-\frac{J_{23}}{12}&
0\\
%%%%%%%%%%%%%%%%%%%%%%%%%%%%%%%%%%%%%%
-\frac{J_{23}}{2\sqrt{3}}&
&
\multicolumn{1}{c|}{\frac{J_{23}}{4\sqrt{3}}}&
-\frac{J_{23}}{3}&
0&
\frac{E_z-3J+\widetilde{E}_{z,2}}{2}-\frac{5J_{23}}{12}\\
%%%%%%%%%%%%%%%%%%%%%%%%%%%%%%%%%%%%%%
\end{array}\right).
\label{eq:HEff_3}
\end{align}
\end{widetext}
\eref{eq:HEff_3} contains two subspaces of virtually identical energies, as marked by the borders in the matrix. All the terms that couple these subspaces can be neglected.

After neglecting the block off diagonal entries in \eref{eq:HEff_3}, also the time evolutions of $E_z$ and $J$ factor because they commute with the remaining entries. The time evolution through $\frac{J}{4}[
\left(\bm{\sigma_3}\cdot\bm{\sigma_4}-\bm{1}\right)+
\left(\bm{\sigma_4}\cdot\bm{\sigma_5}-\bm{1}\right)]\simeq
\frac{J}{2}
\left(\op{0^{\text{R}}}{0^{\text{R}}}-\op{1^{\text{R}}}{1^{\text{R}}}\right)
$ causes only single-qubit time evolutions of the triple-QD qubit, and $E_z$ causes global phase evolutions. The remaining time evolution is
\begin{align}
\label{eq:HEff_4}
\mathcal{U}_{\phi,\psi}^{\text{C}_1}\approx&e^{-i2\pi\left(
\phi m_1+
\frac{\psi}{2} m_2\right)},\\\nonumber
m_1=&-\text{diag}\left\{\frac{1}{4},\frac{1}{4},\frac{1}{4},\left(\begin{array}{ccc}
\frac{5}{12}&0&\frac{1}{3}\\
0&\frac{1}{12}&0\\
\frac{1}{3}&0&\frac{5}{12}
\end{array}\right)\right\},\\
m_2=&\text{diag}\left\{-1,1,1,-1,1,1\right\},
\nonumber
\end{align}
with $\phi=J_{23}/h$ and $\psi=\widetilde{E}_{z,2}/h$.

\eref{eq:HEff_4} causes no leakage for $\frac{1}{3}\sqrt{4\phi^2+9\psi^2}=2\mathbb{Z}+1$, and an entangling operation is realized for $\frac{1}{6}\left(2\phi-3\psi\right)=\mathbb{Z}$. Alternatively, it is also possible to use $\frac{1}{3}\sqrt{4\phi^2+9\psi^2}=2\mathbb{Z}$ and $\frac{1}{6}\left(2\phi-3\psi\right)=\mathbb{Z}+\frac{1}{2}$. E.g., the entangling operation $\mathcal{U}_{3/(2\sqrt{2}),1/\sqrt{2}}^{\text{C}_1}$ gives a CPHASE in the basis $\left\{\ket{0^{\text{L}}0^{\text{R}}},\ket{0^{\text{L}}1^{\text{R}}},\ket{1^{\text{L}}0^{\text{R}}},\ket{1^{\text{L}}1^{\text{R}}}\right\}$ using [see \fref{fig:03}(b)]:
\begin{align}
\label{eq:CPHASE_C1}
\text{Z}^{\text{L}}_{1/\sqrt{2}}
\text{Z}^{\text{R}}_{\left(4+\sqrt{2}\right)/8}
\mathcal{U}_{3/(2\sqrt{2}),1/\sqrt{2}}^{\text{C}_1}=
e^{i\pi\frac{\sqrt{2}-3}{2}}
\text{CPHASE}.
\end{align}

Note that in the construction of \eref{eq:CPHASE_C1}, it was assumed that $J_{12}$ is turned to zero during the entangling operation. Small values of $J_{12}$ can only be tolerated if they are much smaller than $\widetilde{E}_{z,2}$. An alternative gate can be constructed for large $J_{12}$. In this case, \eref{eq:Ham_STQRX} is modified to
\begin{align}
\nonumber
\mathcal{H}^{\text{C}_2}=&
\frac{J_{12}}{4}\left(\bm{\sigma_1}\cdot\bm{\sigma_2}-\bm{1}\right)
+\frac{J}{4}\left[
\left(\bm{\sigma_3}\cdot\bm{\sigma_4}-\bm{1}\right)+
\left(\bm{\sigma_4}\cdot\bm{\sigma_5}-\bm{1}\right)\right]\\\label{eq:Ham_STQRX_2}
&+\frac{\Sigma E_z}{2}\left(\sigma_{z,1}+\sigma_{z,2}\right)+\frac{J_{23}}{4}\left(\bm{\sigma}_2\cdot\bm{\sigma}_3-\bm{1}\right)\\\nonumber
&+\frac{E_z}{2}\left(\sigma_{z,1}+\sigma_{z,2}+\sigma_{z,3}+\sigma_{z,4}+\sigma_{z,5}\right).
\end{align}
\eref{eq:Ham_STQRX_2} contains the exchange interactions $J_{12}$, $J_{23}$, and $J$. Additionally to a global magnetic field $E_z$, the sum of the magnetic field variations at QD$_1$ and QD$_2$ are important $\Sigma E_z=(\widetilde{E}_{z,1}+\widetilde{E}_{z,2})/2$. The magnetic field difference $\Delta E_z=(\widetilde{E}_{z,1}-\widetilde{E}_{z,2})/2$ can be neglected if it is much smaller than $J_{12}$.
Using the equivalent arguments as before for $E_z,J_{12},J\gg \Sigma E_z,J_{23}$, the qubit time evolution is restricted to the subspace $\left\{\ket{T_00^{\text{R}}},\ket{m_1},\ket{T_01^{\text{R}}},\ket{m_2},\ket{S_00^{\text{R}}},\ket{S_01^{\text{R}}}
\right\}$. Projecting \eref{eq:Ham_STQRX_2} to this basis gives

\begin{widetext}
\begin{align}
\mathcal{H}^{\text{C}_2}\approx\left(\begin{array}{cccccc}
%%%%%%%%%%%%%%%%%%%%%%%%%%%%%%%%%%%%%%
\frac{E_z-J}{2}-\frac{J_{23}}{4}&
\multicolumn{1}{c|}{0}&
&
-\frac{J_{23}}{2\sqrt{6}}&
&
-\frac{J_{23}}{4\sqrt{3}}\\
%%%%%%%%%%%%%%%%%%%%%%%%%%%%%%%%%%%%%%
0&
\multicolumn{1}{c|}{\frac{E_z-J}{2}+\Sigma E_z-\frac{J_{23}}{4}}&
\frac{J_{23}}{2\sqrt{6}}&
\frac{J_{23}}{4\sqrt{3}}&
&
\frac{J_{23}}{2\sqrt{6}}\\\cline{1-4}
%%%%%%%%%%%%%%%%%%%%%%%%%%%%%%%%%%%%%%
&
\frac{J_{23}}{2\sqrt{6}}&
\multicolumn{1}{|c}{\frac{E_z-3J}{2}-\frac{J_{23}}{4}}&
\multicolumn{1}{c|}{-\frac{J_{23}}{3\sqrt{2}}}&
-\frac{J_{23}}{4\sqrt{3}}&
-\frac{J_{23}}{6}\\
%%%%%%%%%%%%%%%%%%%%%%%%%%%%%%%%%%%%%%
-\frac{J_{23}}{2\sqrt{6}}&
\frac{J_{23}}{4\sqrt{3}}&
\multicolumn{1}{|c}{-\frac{J_{23}}{3\sqrt{2}}}&
\multicolumn{1}{c|}{\frac{E_z-3J}{2}+\Sigma E_z-\frac{5J_{23}}{12}}&
-\frac{J_{23}}{2\sqrt{6}}&
-\frac{J_{23}}{3\sqrt{2}}\\\cline{3-5}
%%%%%%%%%%%%%%%%%%%%%%%%%%%%%%%%%%%%%%
&
&
-\frac{J_{23}}{4\sqrt{3}}&
-\frac{J_{23}}{2\sqrt{6}}&
\multicolumn{1}{|c|}{\frac{E_z-J}{2}-J_{12}-\frac{J_{23}}{4}}&
\\\cline{5-6}
%%%%%%%%%%%%%%%%%%%%%%%%%%%%%%%%%%%%%%
-\frac{J_{23}}{4\sqrt{3}}&
\frac{J_{23}}{2\sqrt{6}}&
-\frac{J_{23}}{6}&
-\frac{J_{23}}{3\sqrt{2}}&
&
\multicolumn{1}{|c}{\frac{E_z-3J}{2}-J_{12}-\frac{J_{23}}{4}}\\
%%%%%%%%%%%%%%%%%%%%%%%%%%%%%%%%%%%%%%
\end{array}\right).
\label{eq:HEff_5}
\end{align}
\end{widetext}

All the terms in \eref{eq:HEff_5} outside of the marked subspaces are neglected for $E_z,J_{12},J\gg\Sigma E_z,J_{23}$. Neglecting the contributions of $E_z$, $J$, and $J_{12}$ (again these terms commute with the remaining entries in \eref{eq:HEff_5}, and they cause either global phase evolutions, or single-qubit time evolutions) the effective time evolution is

\begin{align}
\label{eq:HEff_6}
\mathcal{U}^{\mathcal{C}_2}_{\phi,\psi}&\approx e^{-i2\pi\left(\phi m_1+\psi m_2\right)},\\
\nonumber
m_1&=-\text{diag}\left\{
\left(\begin{array}{cc}
\frac{1}{4}&0\\
0&\frac{1}{4}
\end{array}\right),
\left(\begin{array}{cc}
\frac{1}{4}&\frac{1}{3\sqrt{2}}\\
\frac{1}{3\sqrt{2}}&\frac{5}{12}
\end{array}\right),
\frac{1}{4},
\frac{1}{4}\right\},\\
\nonumber
m_2&=\text{diag}\left\{0,1,0,1,0,0\right\},
\end{align}
with $\phi=J_{23}/h$ and $\psi=\Sigma E_z/h$. The contributions of $J_{12}$ and $J$ are irrelevant in \eref{eq:HEff_5} because only single-qubit time evolutions are generated: $\frac{J_{12}}{4}\left(\bm{\sigma}_1\cdot\bm{\sigma}_2-\bm{1}\right)\simeq \frac{J_{12}}{2}\left(\op{0^{\text{L}}}{0^{\text{L}}}-\op{1^{\text{L}}}{1^{\text{L}}}\right)$ and $\frac{J}{4}\left[
\left(\bm{\sigma_3}\cdot\bm{\sigma_4}-\bm{1}\right)+
\left(\bm{\sigma_4}\cdot\bm{\sigma_5}-\bm{1}\right)\right]\simeq
\frac{J}{2}
(\op{0^{\text{R}}}{0^{\text{R}}}-\op{1^{\text{R}}}{1^{\text{R}}})
$. Also the phase evolution through $E_z$ is neglected.

The time evolution in \eref{eq:HEff_6} causes no leakage for $\frac{1}{2}\sqrt{\phi^2-\frac{4\phi\psi}{3}+4\psi^2}=2\mathbb{Z}+1$, and an entangling operation is realized for $\frac{1}{12}\left(\phi-6\psi\right)=\mathbb{Z}$. Alternatively, it is also possible to use $\frac{1}{2}\sqrt{\phi^2-\frac{4\phi\psi}{3}+4\psi^2}=2\mathbb{Z}$ and $\frac{1}{12}\left(\phi-6\psi\right)=\mathbb{Z}+\frac{1}{2}$. For example, the entangling operation $\mathcal{U}_{3/\sqrt{2},1/(2\sqrt{2})}^{\text{C}_2}$ gives a CPHASE gate in the basis $\left\{\ket{0^{\text{L}}0^{\text{R}}},\ket{0^{\text{L}}1^{\text{R}}},\ket{1^{\text{L}}0^{\text{R}}},\ket{1^{\text{L}}1^{\text{R}}}
\right\}$ using [see \fref{fig:03}(c)]:
\begin{align}
\label{eq:CPHASE_C2}
\text{H}^{\text{L}}
\text{Z}^{\text{R}}_{1/2}
\mathcal{U}_{3/\sqrt{2},1/(2\sqrt{2})}^{\text{C}_2}
\text{H}^{\text{L}}=
e^{i\pi\frac{3(\sqrt{2}-2)}{4}}
\text{CPHASE},
\end{align}
where $\text{H}$ is the Hadamard gate.

%------------------------------------------------------------------------------------
%------------------------------------------------------------------------------------
%------------------------------------------------------------------------------------
\section{\label{sec:Discussion}
Discussion and Conclusion}

\begin{figure}
\includegraphics[width=0.49\textwidth]{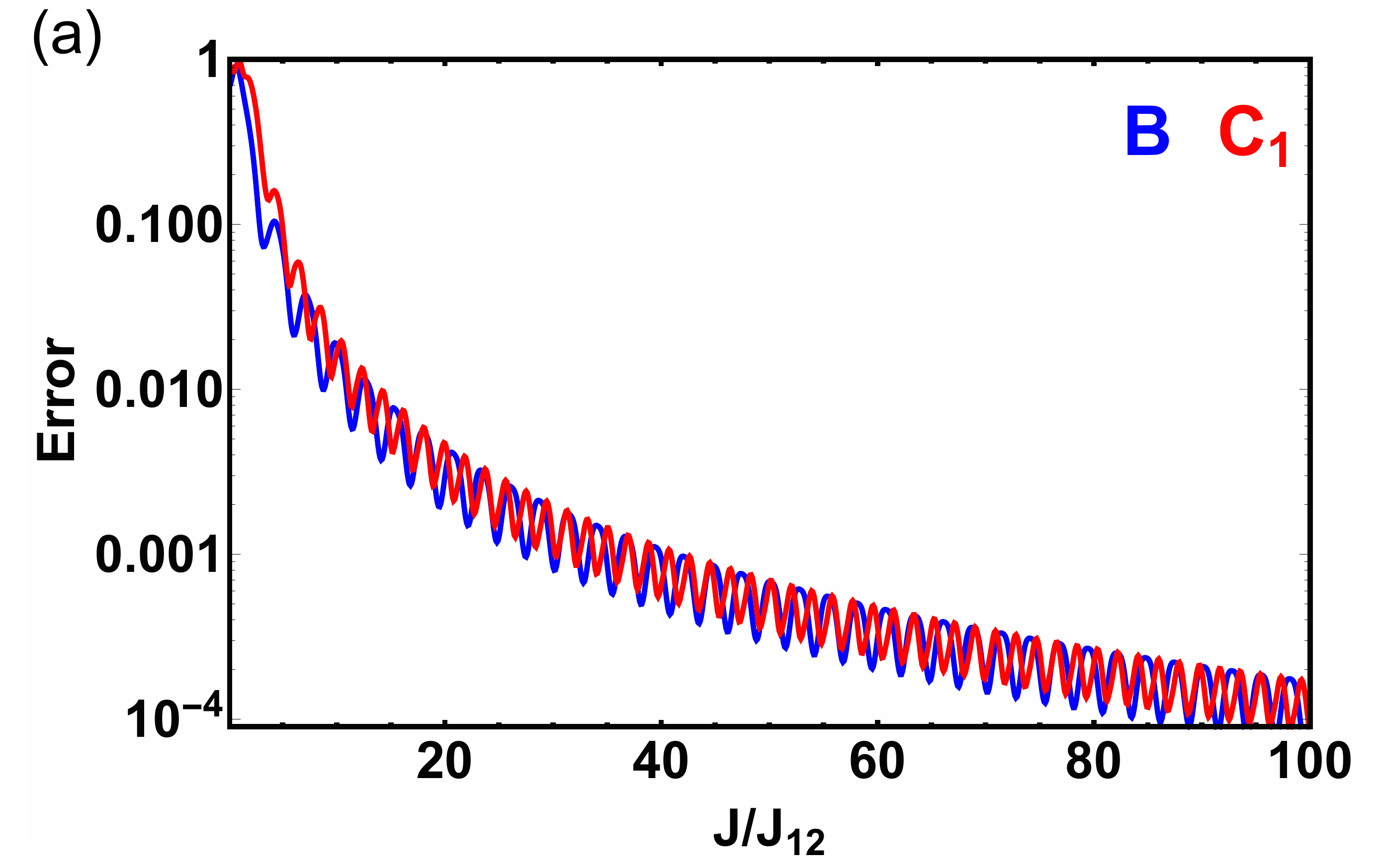}
\includegraphics[width=0.49\textwidth]{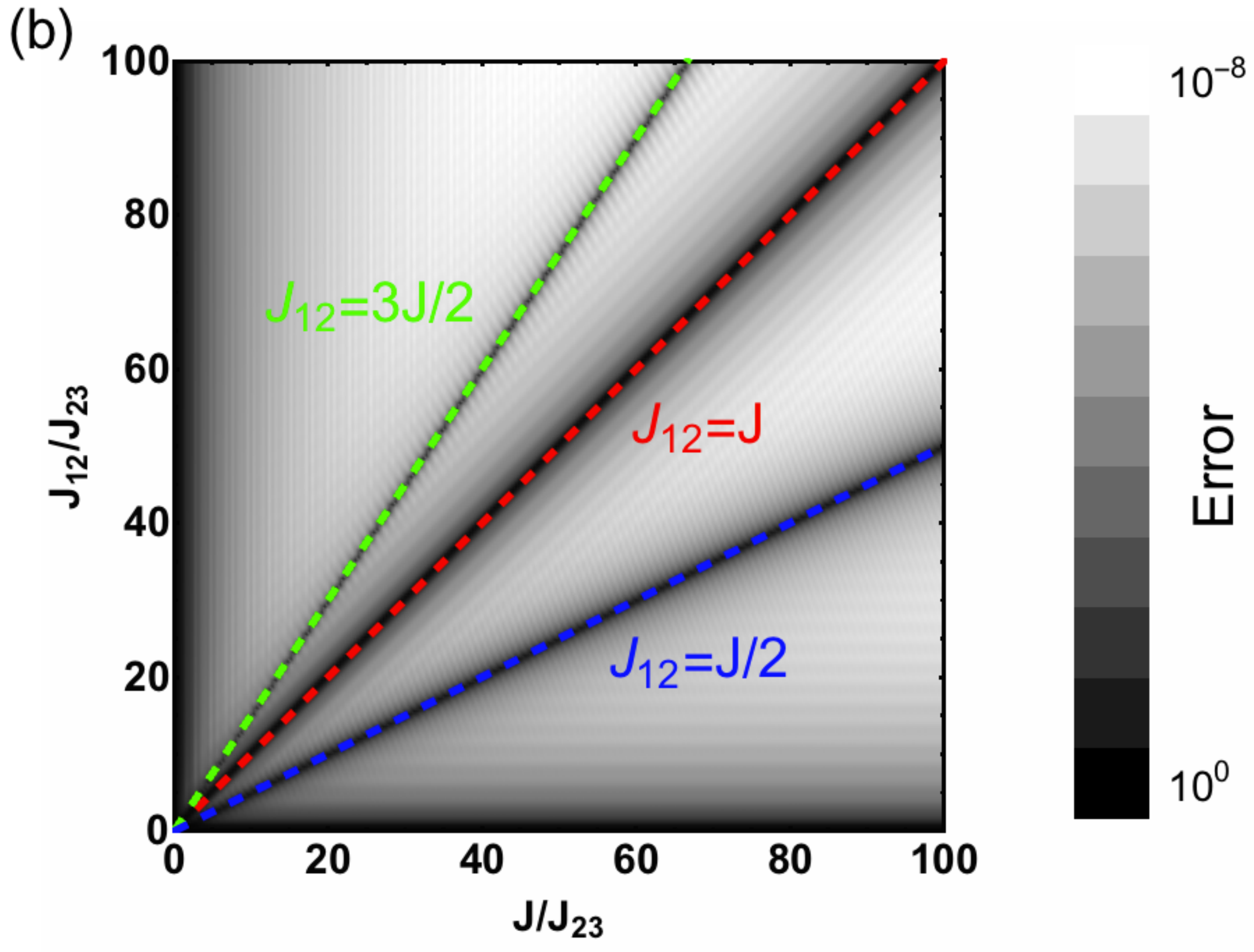}
\caption{\label{fig:04}
Gate errors for the operation sequences of \eref{eq:CPHASE_B}, \eref{eq:CPHASE_C1}, and \eref{eq:CPHASE_C2}. Only the operations $\mathcal{U}_{3/4}^{\text{B}}$, $\mathcal{U}_{3/(2\sqrt{2}),1/\sqrt{2}}^{\text{C}_1}$, and $\mathcal{U}_{3/\sqrt{2},1/(2\sqrt{2})}^{\text{C}_2}$ are analyzed. The gate errors are characterized by the deviation of the entanglement fidelity $F=\text{tr}\left(\rho^{\text{RS}}\mathcal{U}^{-1}_{\text{ideal}}\mathcal{U}_{\text{real}}\rho^{\text{RS}}\mathcal{U}_{\text{real}}^{-1}\mathcal{U}_{\text{ideal}}\right)$ from 1. $\rho^{\text{RS}}=\op{\text{RS}}{\text{RS}}$ is the maximally entangled state of two identical subspaces R and S, e.g., $\ket{\text{RS}}\propto\ket{0000}+\ket{0110}+\ket{1001}+\ket{1111}$, and the time evolutions $\mathcal{U}_{\text{ideal}}$ and $\mathcal{U}_{\text{real}}$ act only on S while R remains unchanged. (a) For $\mathcal{U}_{3/4}^{\text{B}}$ (blue curve) and $\mathcal{U}_{3/(2\sqrt{2}),1/\sqrt{2}}^{\text{C}_1}$ (red curve) the exchange interaction of the exchange-only qubit $J$ should be by more than one order of magnitude larger than $J_{12}$ to reduce the gate error below 1\%. (b) For $\mathcal{U}_{3/\sqrt{2},1/(2\sqrt{2})}^{\text{C}_2}$, $J_{12}$ and $J$ should be large. The gate errors increase for $J_{12}=3J/2$, $J_{12}=J$, and $J_{12}=J/2$ (dashed lines) because of degeneracies in the level spectrum.}
\end{figure}

It has been shown that the exchange interaction can be used to entangle a pair of QD qubits for all the distinct qubit encodings. Besides the single-qubit control, which has been experimentally realized for all the described spin qubits, only exchange interactions between a pair of QDs of different QD qubits are needed. With the flexibility of the spin qubit setup, i.e. by keeping constant exchange interactions (for the STQ or the exchange-only qubit) or allowing local magnetic field variations (for the STQ), very short operation sequences can be constructed to entangle QD qubits. To entangle a STQ with a single-spin qubit or an exchange-only qubit, only one exchange interaction is needed between QDs of the different qubit types. To entangle a single-spin qubit and an exchange-only qubit, a sequence of two inter-qubit exchange interactions is needed.

The advantage of exchange-based entangling operations is the controllability of the interaction mechanism. The exchange interaction depends on the tunnel coupling between distant QDs and their chemical potentials. It has been shown that exchange interactions can be tuned rapidly.\cite{petta2005} The limitations of the proposed entangling operations are similar to existing gate schemes. Local magnetic\cite{burkard1999} and electric field\cite{hu2006} fluctuations are present in semiconductors. Both mechanism cause low-frequency fluctuations of the QD parameters. It is possible to reduce the influence of low-frequency fluctuations by refocusing protocols \cite{bluhm2011,medford2012}, which can also be optimized numerically \cite{cerfontaine2014}. Spin-orbit interactions are weak in typical QD materials like, e.g., GaAs or Si \cite{hanson2007-2,zwanenburg2013}, and they should have minor influence on the proposed operation sequences.

The constructions of the entangling operations in \sref{sec:SSRX} and \sref{sec:STQRX} used a few approximations. It was assumed that the exchange interaction between the QDs of the STQ, or between the QDs of the exchange-only qubit are constantly turned on, while their magnitudes are much larger than the exchange interaction between the neighboring QDs of the different qubits. \fref{fig:04} show that about one order of magnitude difference in the interaction strength is sufficient to reduce the effective gate errors below 1\%. These gate errors are sufficient for quantum computation with standard quantum error correction protocols.\cite{fowler2009,fowler2012,jones2012}

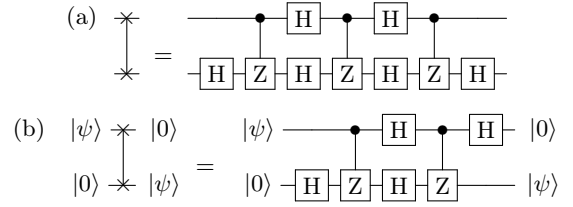
\begin{figure}
\[
\Qcircuit @C=.5em @R=1em @!R {
\push{\text{(a)}\rule{0.2em}{0em}}&&\qswap\qwx[1]&\qw&\push{\ \ \ }&&\qw&\ctrl{1}&\gate{\text{H}}&\ctrl{1}&\gate{\text{H}}&\ctrl{1}&\qw&\qw\\
&&\qswap&\qw&\ustick{=}&&\gate{\text{H}}&\gate{\text{Z}}&\gate{\text{H}}&\gate{\text{Z}}&\gate{\text{H}}&\gate{\text{Z}}&\gate{\text{H}}&\qw
}
\]
\[
\Qcircuit @C=.5em @R=1em @!R {
\push{\text{(b)}\rule{0.em}{0em}}&&\push{\ket{\psi}\ }&\qswap\qwx[1]&\qw&\push{\ket{0}}&\push{\ \ \ }&&\push{\ket{\psi}\ }&
\qw&\ctrl{1}&\gate{\text{H}}&\ctrl{1}&\gate{\text{H}}&\qw&\push{\ket{0}}\\
&&\push{\ket{0}\ }&\qswap&\qw&\push{\ket{\psi}}&\ustick{=}&&
\push{\ket{0}\ }&
\gate{\text{H}}&\gate{\text{Z}}&\gate{\text{H}}&\gate{\text{Z}}&\qw&\qw&\push{\ket{\psi}}
}
\]
\caption{\label{fig:05}
Gate operations to interchange qubits using CPHASE gates. (a) The unconditioned SWAP operation requires three CPHASE gates together with Hadamard gates (H). (b) A simpler SWAP sequence can be realized if one of the qubits is initialized to a fixed state, e.g., $\ket{0}$. Then the SWAP operation with an arbitrary state $\ket{\psi}$ requires only two CPHASE gates.
}
\end{figure}

Besides entangling different kinds of spin qubits, it might also be useful to interchange quantum information between them. \fref{fig:05} shows operation sequences for SWAP operations that only rely on CPHASE and Hadamard gates (cf. \rcite{nielsen2000}). An unconditioned SWAP is realized using three CPHASE gates; only two CPHASE gates are needed if the state of a qubit should be transferred to another qubit that is initially in $\ket{0}$.

Altogether, very efficient operation sequences have been constructed to couple and interconvert different kinds of spin qubits. These operation sequences can couple all the standard qubit encodings in one, two, and three singly occupied QDs. Only the established single-qubit manipulation protocols are needed that have been successfully realized for all the qubit encodings. Different qubits are coupled using exchange interactions that are well controlled experimentally. With the current efforts to build larger arrays of tunnel-coupled QDs,\cite{takakura2014,delbecq2014} the proposed operation sequences can be tested directly. The interconversion of different spin qubits allows to use all the advantages of the different QD setups in large arrays of QDs. E.g., it is known that few-electron qubits couple stronger to cavities\cite{taylor2006-2,burkard2006,taylor2013} or metallic gates\cite{trifunovic2012}, while single-spin qubits have extremely long coherence times.\cite{balasubramanian2009,pla2012} Therefore the described operation sequences are another useful ingredient on the way towards quantum computation with large QD networks.

%------------------------------------------------------------------------------------
%------------------------------------------------------------------------------------
%------------------------------------------------------------------------------------
\begin{acknowledgements}
We are grateful for support from the Alexander von Humboldt foundation.
\end{acknowledgements}

%------------------------------------------------------------------------------------
%------------------------------------------------------------------------------------
%------------------------------------------------------------------------------------
\bibliography{library}
\end{document}